# About a possible path towards the reverse engineering of quantum mechanics.

*Alberto Ottolenghi*[1]

*Between London and Paris.*
*Between 9 March and 22 September 2011.*

**Abstract.**

An out of the box intellectual path exploring the foundations of quantum mechanics is discussed in some detail, in order to clarify why a possibly different way to look at the relevant fundamental questions can be identified and can support further research. Two key concepts arise. (1) Einstein critics to quantum mechanics could be taken seriously, but ironically, in order to really do so, one would have to take seriously also some of Lorentz critics to special relativity - both in a possibly more modern way; such interconnection possibly having been a blocking factor to openly discuss some of the cross implications of alternative views about quantum mechanics to date. (2) The probabilistic interpretation is a by-product of (a) quantum evolution equations, (b) conservation laws for the combination of measuring system and measured object and (c) persistency of calibration of the measuring system – as such there is no intellectual conflict whatsoever between hidden variables theories and probabilistic interpretation, provided we consider multicomponent hidden variable models and we allow for the existence of an underlying network. The implications of such concepts, in particular for the development of a microscopic quantisation program, are heuristically discussed or preliminarily explored. The interaction between science, philosophy and religion is at times discussed in order to show how the limitations of each might have mixed up concepts which could be more easily untangled today, allowing then to look for a strong constructive cross feedback between knowledge, which is incomplete by definition, and belief, which can be often understood *a posteriori* as not fully digested knowledge.

*When the world*
*was substance without form*
*and form without substance*
*(Genesis, 1, 2)*[2]

**Background.**

The long lasting debate about the foundations of quantum mechanics is of paramount relevance to address the raising need for a more effective and comprehensive framework allowing the transmission of scientific knowledge across generations. Such need is rising with exponentially diversified specific knowledge and cross links among fields, technologies and concepts requiring a simple and intuitive cross mapping, particularly once quantum effects come into play. The need for simpler intuition is particularly strong in the domain of experimental work and search for applications. In such fields, day to day experiment planning needs have shown that personal intuition brought to the quantum mechanical realm is actually much more of a help than one would expect and that is most effective when it is not just based on correspondence principles, but also on a personal tentative deeper physical understanding of what might be underlying quantum physics concepts. It is then far from surprising that Bell's work originated close to experiment or that often experimental quantum chemists and engineers seem to have been more creative and successful in their intuitive understanding of the quantum world, being less often restrained by the more formal part of the physical theory – as e.g. in the case of the inventions of NMR, its optical analogue and many of their key applications. Furthermore, with an increasingly expensive experimental frontier science, higher intuition is key in order to allow more effective decision-making about research strategy choices and their risk-reward profile.

---

[1] Unaffiliated: private research. Contactable by email: alberto.ottolenghi@hotmail.com

[2] Translation based on the notes to the commentary of the Genesis by E.Munk, 1981, FSOL, Paris. An ideogram based translation of the Hebrew words Tohu and Vohu would be: "formal sign of immaterial peg" and "content of immaterial peg". If a translator wanted then to promote a more high energy physics modern view of the Genesis, "string" rather than "peg" would be used for the Hebrew letter Vav which becomes mute to support the "u" at the end of each word. Whatever the approach to the etymology of the words, there is here an intuitive understanding that lower energy matter and fields are somehow a synthesis of formal sign and content of some sort (just like writing and other forms of graphical communication), which would have existed as distinguished entities in some way at the beginning of the formation of the universe. This is intuitively in line with many other forms of more or less sophisticated beliefs, including possibly some views of platonic philosophy and, as some might say, with the modern neo-platonic cobordism approach to a unified forces model of the world in some high dimensional theories.

This paper is written to record my personal intellectual path on the matter, tracing its seed ideas from mathematical physics and mathematics, then matured through experimental physics work and later taking some conceptual shape during a long period of almost total isolation from the broader scientific community. There was no intention on the path to look in this direction, just a call for understanding I could not stay away from. Nor is there any intention here to state: "this is the right approach". Rather this paper is written to highlight that there might be other ways to combine and mix old and new ideas about quantum mechanics, which could lead to different decision making processes in some research projects or fields. If there is a base of higher understanding here, then I might be contributing to the effort of those who seek a Shumpeter[3] *creative destruction* event for some concepts in physics. Else it will just be my personal search. Even if far from being perfect, I believe it is now time to test my reasoning openly, including in order to see if research work on these lines could be funded, by private or by public institutions, and/or collaborations with others could be discussed. This paper is then focused on ideas, how they came about and how they are related to other ideas. With hindsight, it seems also a good story to tell.

**Executive summary.**

There seems to be a number of supporting elements suggesting that it should be possible for quantum mechanics as a mathematical theory to be reversed engineered in the sense of hidden variables processes. Any relevant macroscopic implication of such underlying processes, not simply taken into account by quantum mechanics as an effective theory, would then be captured either by taking the semiclassical limit, typically with the appearance of a Maslov index, or by introducing quantised fields (when viable), both of which would allow to capture that part of the missing information which is relevant for the problem at hand. Such a framework would be of mathematical relevance as one is effectively seeking proof of a purely mathematical conjecture broadly as follows: *for every pseudo-differential Schrödinger or Dirac operator, including in curved spacetime, there is a multi component diffusion process within some mathematically well defined network for which the evolution equation generated by the operator is a well defined thermo-dynamical limit*. It should be clear nevertheless that, even if a proof of a conjecture of this kind could actually be built, and possibly allow to extend the concept of pseudo-differential operators, this still does not mean that the processes which could be identified through such formal reverse engineering are actually physical processes: that is for experiment to test. If formally sound, such an approach would provide a procedure allowing a quantisation "from below" (a microscopic quantisation) as opposed to the introduction of operators or functors, which would be a quantisation "from above" (an *a-posteriori* formal, effective quantisation).

Historically such an approach "from below" has been considered impossible in principle. Nevertheless, 20 years ago, G.N.Ord has explicitly built a diffusion process in one dimension for a single particle. Therefore it is not necessary anymore to discuss conceptual or logical implications, but there is an actual formal mathematical process to manipulate. Ord framework less obviously implies that in order to achieve a reverse engineering of quantum mechanics one needs also some reverse engineering of special relativity and, in addition, a number of speculative assumptions about the dynamics and fundamental entities which take part in the identified processes. Because of the current knowledge from the study of macroscopic quantum phenomena, such as superconductivity, lasers, quantum Hall effects, macroscopic spin and charge density waves and so on, the most natural assumptions can be chosen with some analogy to electron instabilities and low dimensional condensed matter physics. Whatever such assumptions one might choose, they seem to have waterfall implications on the modelling of the source of the gravitation force, the source of inertia and so many related topics - therefore likely just to raise much scepticism. This has prevented me from discussing ideas on paper until now, hoping rather to achieve a first result from a modelling point of view. But until now I have not had the time and peace of mind required to try to build such a prototype model, while I believe that the underlying ideas should be more openly tested. Among the many reasons, also because they seem to lead to a microscopic description of Planck-like scale physics closer conceptually to genomics or some soft matter systems, rather than esoteric physical concepts trying to interpret more abstract or complex mathematics. This in turn does suggest that there might be some solid common sense somewhere in the approach and that there could be a natural and transparent hierarchical structure of nature all through from the very micro to the very macro level. Which in turn would support, maybe reversely, the old Bohr ideas about quantum and life - which also did motivate e.g. Gamow interest in the geometrisation of DNA-RNA coding transmission - including that quantum coherence has a role in the stability and formation of leaving beings, as more recent research work starts to tentatively re-explore and formalise.

Rather than a universe as a "leaving being", which is a popular idea among the neo-pagan inspiration of part of modern science and ecology, one would face a framework whereby both the universe and leaving beings are different higher hierarchical structures reflecting the ultra small microscopic structure of matter. Such a *Weltanschauung* is mainly a consequence of the informed assumptions that: (i) quasi one dimensional objects are the underlying entities of quantum processes; and (ii) to get a credible hidden variables picture one has to consider also special relativity as an effective theory. The conceptual tools which derive from the ideas here discussed might also allow to easily bridge between quite different mathematical tentative fundamental descriptions of the microstructure of nature, such as string theory, non commutative geometry unification theories, loop quantum gravity or scale relativity. Furthermore, where the key considerations made here correct in some way, there should be a broader class of models based on multi component diffusion processes within a network, which should include cases of relevance for other areas of knowledge, outside of quantum mechanics, but displaying some partial quantum-like features. In this sense, liquid crystals are maybe an already known case, whereby both the network and the diffusion are provided by different degrees of freedom of the same physical entities. A further brainstorming is then also included, in order to shortly discuss some possible implications and some subjects where similar modelling ideas could eventually be considered.

---

[3] I am referring here to the idea of the first half of the twentieth century Austrian born banker and American naturalized economist - who went through banking and personal bankruptcy - about the nature of more extreme transformation within economic, industrial or business life cycles: effectively the conceptually deeper, economic or business equivalent of singularities "catastrophe" theory.

**The intellectual path: a not fully random walk of ideas.**

*1) What is supersymmetry in classical mechanics?*

My first research project was about showing, by constructive calculation and upper bound estimates, that a small internal local dissipation which was retained between each couple of degrees of freedom of a classical mechanical system, as represented by an imaginary part of the relevant frequency, would smooth the effect of perturbations and bring back the system in some way to a fully integrable systems with globally preserved conservation laws, avoiding the singularities of the global solutions which would else arise and which represent macroscopic build up of dissipation when particles fill the configuration space of the system. The identified framework, never fully explored to my knowledge, aimed at showing that, if the dissipation caused by global solutions is fully recycled within the system, then the mechanical system can in principle self preserve globally its original conservation laws also in the presence of a continuum of particles. Therefore there can be globally conserved quantities in presence of dissipation, not dissimilar from soliton solutions arising out of a mix of dissipation, dispersion or other nonlinearities, but here for a classical Hamiltonian finite degrees of freedom dynamical system and modelled by the interplay between real and imaginary components of some core frequencies. A resummation of classical perturbation series by extending the domain of some variable is intriguing also because it represents a regularisation of the zero order approximation of semiclassical quantisation formulas for quasi-integrable systems, therefore a first step on the way towards the regularisation by resummation of perturbation singularities in the semiclassical limit. In light of this connection to quantum systems, the use of complex frequencies is quite natural for a Hamiltonian system and hints to a meaning of quantum equations in the semiclassical limit.

An abstract non-constructive proof in two degrees of freedom did already exist but this was overlooked by myself and by both my old and new supervisor. Nevertheless, the natural idea arising from the obtained constructive proof of convergence for two degrees of freedom was that more degrees of freedom could probably be also regularised through the use of additional excess complex variables and/or larger Clifford algebras. Which in turn indicated that there should be a conceptual analogy between extended complex or higher Clifford algebra formalism in classical mechanics and extended supersymmetry in quantum field theory. By reverse analogy, this would suggest that supersymmetry in quantum field theory is probably about attaching to each field an independent field of its collective excitations - for fermion fields, collective excitations as a boson field build through bosonisation (two-fermion field), for boson fields collective excitations as a fermion field build through fermionisation (infinite-boson field). The supersymmetrical partner field would then provide the degrees of freedom that can absorb and release back energy or other conserved quantities in order to avoid destabilisation of the combined system (singularities). But within the current understanding of supersymmetry formalism such partner fields are treated as excitation of independent fields, which would be then an idealisation. If this intuition was right, it would be normal that to observe supersymmetry partners of known particles one needs a collective state to form first locally and therefore a much higher luminosity and lateral size of generating beams, making the search for supersymmetry particles possibly more expensive than one would estimate if taking the literal formal understanding. Whether right or not, there seemed to be here a sign that at the interface between classical and quantum mechanics there could have been still not fully explored concepts of relevance for fundamental physics.

*2) Can we compare quantum mechanics to hydrodynamic potential flows with small viscosity?*

Likely because of a first sight conceptual and possibly formal analogy with the topic above, I was then asked to focus on the search for a possible proof of a full or partial equivalence between (i) solutions of Hamilton-Jacobi evolution partial differential equations with a viscosity term, in the zero viscosity limit, and (ii) solutions of same-Hamiltonian quantum evolution equations in the semiclassical limit. The technique to be used in such a search leveraged on the local dimensional reduction of an action like functional and variational analysis on the remaining variables, as inspired by sympletic capacities variational techniques, which in turn are generalisations of the EBK ideas used to search for periodic orbits originally developed to allow first order semiclassical quantisation of non-integrable systems. The use of local dimensional reduction highlighted that one could look locally at the respective evolution equation as a system of countable variables. Therefore, if the conjecture was right, then some underlying dynamics of countable elements could be used to model the system, both in the zero viscosity limit (already known) and in the semiclassical limit (therefore a multicomponent hidden variable model).

For Hamilton-Jacobi equations with a small viscosity term, a single component diffusion process that models the evolution flow is often known. Therefore a multicomponent process could allow the modelling of a quantum mechanics equation, at least close to the semiclassical limit. As a by-product, one should have not necessarily intuitively expected full equivalence of the two solutions, as extra degrees of freedom should allow more exotic dynamics. Heuristically one would question if such equivalence would break down at least little before the respective limits are taken and close to an unstable fixed point. This is because the quantum tunnel effect can first cause part of the global solution to move from a bounded orbit to an unbounded orbit domain, allowing solution points to move then across the instable fixed point configuration space coordinate to achieve a configuration which would not be allowed for a viscosity term solution and which would be, in principle, topologically not equivalent in some sense.

If equivalence between semiclassical and zero viscosity shocks was the case close to unstable fixed points, then something relevant had to be understood at the limit or in general, something which I could not conceptually identify. On the opposite, if multiple components processes could allow to model quantum mechanics equations, then a lack of full equivalence might be derived from an analysis of the impact of the extra internal degrees of freedom. I then tried to find a case where the two solutions were different and I believed I had identified a counterexample from a time independent Hamiltonian, but it was probably correct only up to just before the limit was taken and not at the limit. Even if that was the case I did expect that a similar time dependent counterexample could be build, possibly more easily through numerical simulation, although I did not pursue this option for very long, possibly for lack of access to the relevant information and know how.

By combining the ideas that (i) there could be a subtle difference of the solutions above caused by additional diffusion components in the quantum case and that (ii) an imaginary part of the frequency could be looked at as a recycling of dissipation within the system allowing anomalous conservation laws, I was then lead to the reading of the Schrödinger evolution equation as an effective theory of an underlying viscosity-like diffusion process, whereby the dissipation is a function of the variations of volumetric density and not of variations of the velocity potential, formally resulting in an imaginary viscosity formalism, which is that of the semiclassical form of the Schrödinger equation. In the semiclassical variables – i.e. phase and square modulus of the wave function (which can be looked at as a proxy of a volumetric density) - the imaginary part of the complex velocity potential is a logarithm of the volumetric density, therefore its variation a percentage variation and its contribution similar to that of an entropy like measure. A Schrödinger equation can be then written as a complex Hamilton-Jacobi equation with small viscosity, whereby the velocity is complex and the viscosity term is pure imaginary with intensity set by the ratio between the Planck constant and the mass. Alternatively one should also look at the Schrödinger equation in the semiclassical variables as a velocity and density field non-linear partial differential equation: a Hamilton Jacobi equation with density dependent "dissipation", which can be both exothermic and endothermic. Such non-linear PDE can then be linearized into the Schrödinger equation ("hidden linearity") through a change of variables to the quantum mechanics complex wave function. Just as the Burgers equation (which is non-linear and is equal to the free particle Hamilton-Jacobi equation with a small viscosity) can be linearized into the diffusion equation though a Hopf-Cole transformation. But even written in the semiclassical variables the Schrödinger equation has no shocks and preserves in a modified form the conservation laws, differently from the Burgers equation. This is due to the form of the density dependent terms ("quantum potential") that avoid both excessive compression (no shocks) and excessive diffusion (no total separation of a single wave packet into two or more separate packets) in a way similar to some soliton systems, but with "hidden" linearity, global energy conservation and lack of local energy conservation.

These considerations highlighted that there might be also a simple fundamental and intuitive reason for the formal parallel between action like generating functions in classical mechanics and thermo-dynamical potentials in thermo-dynamics; between the Hamilton-Jacobi equation in classical mechanics and the equation of state written for the thermo-dynamical potentials. They might be just both equations of state for same thermo-dynamical potential. This point of view might also clarify what was Einstein intuition about particles as non-linear soliton like solution of some unification theory and how did he not seem to have taken into account the "hidden linearity" of quantum mechanics when looked at from a non linear PDE point of view, which is different from e.g. soliton systems.

Looking back, I must say, I was also trying to make more interesting a mathematical problem that was otherwise a bit boring for a physicist! I was also possibly unaware that I had made partially mine the horizontal conceptualisation methods of much of modern pure mathematics, which are less often used in physics where vertical knowledge is so deep to often prevent such approach – which is normal as fundamental physics deals with the physical world, while fundamental mathematics deals with psychology, representation, numbers and language.

*3) A macroscopic quantum phase: can the same source contribute to both quantum fields and gravity?*

It was then natural to search for a class of multi component processes underlying quantum evolution equations. But I could not identify reasonable ingredients and I did not have access to the appropriate people to discuss this. Neither I did know about the first pre-prints by G.N.Ord, just posted at the time in Canada and which came to my attention only many years later. Multi component diffusion modelling is a wide and highly developed area of research, but none of the classes of models I could find seemed to fit the needs of the problem at hand, nor to be adaptable, often because developed too close to specific applications which allowed approximations not applicable at all in my case. I then tried to go back to the drawing board and to check first how my understanding of experiment changed due to such ideas, and second how the broad and stimulating study of macroscopic quantum physics might allow focusing my search - by using with the due precaution the analogy between complex order parameter and wave function and leveraging on the macroscopic intuition of the phenomena, which is lacking for basic quantum systems.

I was then given the opportunity first to join a very good experimental team and take part in photoemission analysis work. Direct experiment remains the most powerful tool available: I had a fuzzy and possible more advanced intuition of quantum processes on my side and I could immediately remark the lower level of intellectual conflicts arising and the shorter time required to get a good understanding of the measures being done or to coordinate thought and action for precision measurements. This provided encouragement that something of the fuzzy intuition at hand was possibly relevant and allowed to focus the chaos of questions arising, as only experiment can do.

Shortly after I was given the privilege of choosing an experimental subject among a list and, also because of the macroscopic quantum mechanical physics, I chose to perform an X-ray diffraction and diffusion study of the temperature dependent structural phase diagram of the synthetic (poor) metal family of monophosphate tungsten bronzes with pentagonal tunnels ("MPTBp"), whereby alternatively orientated quasi-2D planes of antiferroelectric tungsten trioxide receive donor electrons from the phosphate dioxide groups separating and connecting them. The conduction band of tungsten trioxide crystals is empty and built out of three quasi-2D conduction electron bands, which are superposed, one perpendicular to each axis. In MPTBp, due to the diagonal orientation of the tungsten trioxide, the three quasi-2D bands are cut into three quasi-1D bands, which inter-cross to build a quasi-2D structure, which in turn alternates in orientation along the third dimension. The system is to a certain extent a zigzag (twisted) dielectric analogue of the superconducting quasi-2D copper oxides. Or a solid version of quasi 2D liquid crystals.

Intriguing enough, the temperature dependent phase diagram of this family of synthetic metals shows incommensurate and commensurate structural phase transitions at temperatures where the conductivity has some form of anomaly, which are understood to be transitions of the charge density wave type, whereby transition temperatures are quite high and tend to increase above room temperature with decreasing electron concentration, likely boosted by a cooperative interaction between the charge density wave condensation and the antiferroelectric transition of the underlying tungsten trioxide, possibly stabilising the low density charge density waves at higher temperatures and causing or enabling the higher harmonics of some of the modulations – which might be also a sign of polaron-like condensation, a natural possibility in a highly polarisable environment. For MPTBp of lower conduction electron concentrations, diffraction peaks show a four sub-peaks structure suggesting the coexistence of multiple structural states. Such states being possibly of different chirality, as suggested by crude simulations of the form factor of the diffraction pattern, probably in relation to the two, or more, possible circular multi-polar distortions on the tungsten trioxide perovskite structure, which could be induced or stabilised by the alternate orientation of the $WO_3$ octahedra.

This was a perfect topic to challenge and stimulate my fuzzy ideas about foundations of quantum mechanics. For many important cases of electron instability driven macroscopic quantum phases, it seemed that the formation and macroscopic stability of a quantum order parameter was enabled or stabilised through the interaction with an underlying network, was it through electron-phonon interaction or by the role played by the underlying antiferroelectric or antiferromagnetic network. From this point of view, the role of antiferromagnetism in cuprates had to be looked more as a catalyst to enable spin matching for Cooper pair formation, whose cooperative condensation is then affected by the electron-electron or electron-phonon interaction, either cooperatively or in competition to the underlying network dynamics, depending on the level and type of frustration of the underlying antiferromagnetic order and the electron concentration.

But to use this type of intuition for the search of a multi component diffusion process to build a quantum function it was necessary to introduce some form of network, which seemed to be an excessive stretch of the imagination. On the other hand, the framework of electron instabilities provided a hint on how to identify the components of the process, i.e. anything with properties similar to conduction electrons and holes, with up and down spins and couple formation.

But there was also another relevant lesson to learn. In MPTBp high temperature pre-transitional fluctuations provided indication of relevant phonon involvement into the quantum macroscopic charge localisation process, while some of the conductivity measurements suggested that residual low temperature strong electron-electron interaction was present in some of the members of this synthetic metals family. All this suggests that a strong electron-phonon interaction is present in MPTBp at high temperature, possibly connected to the antiferroelectric instability of the underlying $WO_3$ network, which would cause the high critical temperature phase transitions; but due to a cooperative contribution of the anti-ferroelectric phase transition fluctuations, the electron-phonon interaction could be so strong that once the crystal modulation is generated to balance the charge condensation and the conduction bands are modified accordingly, a residual electron-electron interaction is at times still strong enough to be detected, possibly further helped by the lower electron concentration in selected members of the MPTBp family. The formal description of such a transition from a strong electron-phonon interaction into the sum of weak electron-phonon and residual electron-electron interaction through charge condensation induced modulation is obtained through the application of a Lang-Firsov transformation which rewrites the high temperature Hamiltonian on the new lower temperature modulated crystal structure, resulting in new modified electron bands with a weak electron-phonon interaction involving the new modulated phonons and a residual electron-electron interaction (i.e. because phonons are generated by electrical charges movements, at least some of the phonons and electrons degree of freedoms are the result of the same type of fundamental entities and can cross transform[4]). Normally the residual electron-electron term is so small to be negligible, but possibly not in this system at least for some members of the MPTBp family.

I was then studying a solid state system quite likely showing physics which might have had relevant analogies to a high energy field theory where a very strong quantum gravity like interaction is split, through a lower energies condensation of energy into matter, into a weak gravitational force and a residual electroweak field - therefore a graviweak unification framework. It was then interesting to compare by analogy such two different types of physics. Including supporting the intuition that the condensation of energy into matter might have had a role in making gravity the separate weak low energy force that we know. Which in turn suggests that there should be some conceptual simple relationship between any dynamics underlying quantum mechanics, inertia and gravity or unification, just as there seemed to be for MPTBp between charge condensation into macroscopic quantum density waves and the interplay between antiferroelectricity, nonlinear modulation of crystal structure, electron-phonon and electron-electron interaction.

Such a connection for gravity and quantum mechanics foundations would not be surprising if we recall the famous thought experiment invented by Einstein to try to disprove the universality of the Heisenberg principle, but then shown by Bohr that it was not breaking Heisenberg principle thanks only to the gravitational red shift of light, supporting the intuition that gravity's influence on electromagnetism is a contributing microelement of quantum uncertainty - as best clarified by Pais in his book on Einstein. Furthermore this type of picture brings additional intuition into the Ashtekar formalism where a linear form of gravity is obtained in the complex domain while the real section shows non-linearity as if there had been a phase transition from a high energy linear gravity to a low energy non linear gravity through a complex rotation generated by matter condensation; not dissimilar to a crystal modulation arising from charge density wave condensation or to the mathematical description of quasi-crystals as non commensurate sections of a higher dimensional regular crystal structure. Even the speculative discussions in the sixties by Victor Starr in his book about negative viscosity at the interphase between ocean and atmosphere could be then looked at as a search for a descriptive basis of quantum semiclassical macroscopic phenomena, i.e. where the quantum phase is such that the quantum order parameter is only real and in very unstable situations can move to the negative axis.

But despite these possibly high-level links and conceptual ideas at the time, the formal search for a rigorous mathematical route could not even be started until such links and ideas had been more formally defined. Therefore any reverse engineering of even the most simple equation of quantum mechanics seemed just an unreasonable task. The path was only getting more difficult and apparently desperate also because of the diversity of topics interconnected, although the fuzzy ideas were more and more appealing for the same reason.

*4) Probability interpretation from conservation laws plus persistency of calibration.*

Academic opportunities provided to me became geographically quite challenging and opportunities in the business world came up: I then felt the survival need to try the career change. It worked, but I continued to think about foundations of quantum mechanics as something was still ticking in the back of my mind and I could not stay away from it. Much inspired by the speculation of the possibility of micro endothermic shocks in the semiclassical limit, I questioned myself if, in order to get a better insight about the meaning of the quantum wave function, one should revisit conservation laws as applied to the combined system of the measuring instrument and the quantum system on which the measure is performed – as wave function freezing into an eigenstate requires at times to move to a higher energy state in order to comply with quantum rules.

---

[4] This is a general idea applicable in other cases, and which for electron spin and their magnetism is part of the intuitive basis of the SO(5) theory of superconductivity and antiferromagnetism by S.C.Zhang, where electron spins can pair into Cooper pairs or can neutralize antiferromagnetic localized spins. Although such theory requires for additional interaction to be added as if it was between completely separate independent degrees of freedom, it does clarify for magnetic system the idea of network to gas interaction and potential for exchange of degrees of freedom.

If we take the quantum system e.g. in a state which is not an eigenstate of the observable being measured, then conservation laws require that any measure causes an exchange of an amount of the observable quantity between the quantum system and the measuring system. More precisely, as the quantum system moves from a state that is not an eigenstate of the measured observable to a state which is an eigenstate, the quantum system average of the observable being measured increases or decreases through the measure. For a measure of energy then, the measuring process can be both dissipative or increasing the energy of the quantum system, therefore providing circumstantial support to the idea of the existence of endothermic events close to the semiclassical limit[5]. The variation of the observable for the system undergoing the measure must be then matched by opposite sign and equal amount variation for the measuring apparatus. The measuring of an observable is then a process which forces the quantum system to adapt to the measuring instrument through a transition to a configuration which can be stable without any more exchange of the observable to be measured: an eigenstate. If then repeated measures are performed within a short period of time on a very large number of substantially equal quantum systems, the measuring instrument will have to absorb or release a large number of amounts of the observed quantity. If we want then the measuring instrument not to change its macroscopic state, i.e. to remain calibrated, then we need the sum of all such absorbed and released quantities to be approximately zero. For such to be the case, provided the wave function satisfy the equations of quantum mechanics before and after the measure, the distribution of final states for the quantum systems of same initial state being measured has to follow a probability distribution numerically equal to the square modulus of the wave function of each of the substantially equal quantum systems. It can be then understood, that the so-called probabilistic interpretation of the quantum wave function is numerically correct, but conceptually not accurate and skipping a few steps to get to the right final result.

The probabilistic interpretation is then the result of the fact that (i) post measure each quantum state can be stable only by turning out to be an eigenstate through exchange and adaptation to the measuring system, and (ii) to maintain calibration, the observable being measured should not change on average for the measuring system. Therefore persistency of calibration enforces a rigidity that, once taken together with quantum equations evolution dynamics, causes the well-known probability distributions. If the variation of the observed quantity on average is not zero, the instrument looses its calibration (as e.g. misalignment, where "absorption" is micro-movement of its centre of gravity) and the end distribution of the quantum systems is not the one expected from the probabilistic interpretation of quantum mechanics. For the purpose of performing a measure this does not matter much, as for lost calibration the measure is rejected - although the reverse reasoning is already at times used in practice in order to calibrate to a very high precision an experimental setting. Having this clear, the search for a hidden variable dynamics is then not in any sort of contradiction with the probabilistic interpretation of quantum mechanics and the quantum probability mystery should be reduced in principle to an understanding of the underlying dynamics and how can this be used to model the measuring process.

To perform EPR type analysis for example, one then needs to consider different cases. If e.g. the two measuring systems are not just part of a single measuring apparatus and the measures on two parts of a tangled two parts system are done almost at the same time, provided that the two parts of the two particle wave function more concentrated close to each measuring system are almost separated in some way, then each measure can result in uncorrelated amounts; but multiple measures would reproduce the probability of the initial tangled configuration if both measuring systems remain calibrated - other cases will just require discussion on the same lines. Bell work focuses on implications of breakdown of Lorentz invariance due to superluminal signal between the two measuring system, which is not the case in the case just discussed as it is the local calibration of each system which enforces the quantum probability result. Nevertheless, as it will be discussed below, in a multi-component hidden variables framework, Lorentz invariance is lost at a different level and it is not fully clear how it can be recovered, at least effectively, in all relevant cases. From this point of view, EPR and Bell analysis would seem to be identifying and discussing the right difficulties in some way, but not necessarily to be looking at them with the right goggles.

As the measuring apparatus becomes smaller and of size more comparable to the object being measured, the effectiveness of such freezing process becomes lower, therefore in mesoscopic systems one naturally expects this point of view to become more useful in planning and interpreting experiments as it gives a sense of the intermediate dynamics. In case of weak measurement, in the sense of Aharonov, there is no exchange and the measured system is not perturbed. Nevertheless from this description of the measuring process, one would expect in principle weak measurement to be possible only on zero mass objects and only very approximately on massive particles, provided the measuring system is of mesoscopic size.

*5) 1D single particle Ord quantum mechanics: what are the underlying processes?*

Shortly before a gardening leave between jobs I was kindly referred to the arXiv website by Cornell, the existence of which I did not know yet, to read about the measures of the acceleration of the universe expansion rate and about much of the new astronomy and astrophysics developed on the back of all the ambitious satellite and outer space missions of the years just before and after the start of the new century. On arXiv I could find G.N.Ord preprints, ironically dated very close to when I had first looked at multicomponent dynamics.

---

[5] A quantum measure on multiple copies of a sample system, as here described, and taking the semiclassical limit on a global solution (therefore a continuum of particles), both do force an exchange with an external entity of energy or other observable quantities.

Here one could find a first step towards the identification of a quantum diffusion process, with a full local model in 1D for a single particle. The fact that a multicomponent diffusion process was used by Ord was exactly in line with my original intuition. The fact that the diffusion particles would need to be moving at the speed of light and that the diffusion was taking place within a rigid network was further supporting the idea of a link to inertia and gravity. Mass could be then some form of localisation of light-like entities through the action of the underlying network, whereby gravity would arise as elasticity-like or rather spin-nematic-like waves transferred through a not completely rigid network, including when generated by quantum diffusion – provided such an underlying network could be a physically acceptable concept of some sort. Ord used a fully rigid point like network without structure and a (quasi-complex) rotation like dynamics among two components in the process, using also the two direction of motion as component indexes, therefore fictitiously obtaining a four components diffusion process, which allows in the continuum limit to generate a quasi-complex structure in 1D with the wave mechanics behaviour. This might also have limited the success of the model, as a rotation like dynamics using the two directions of motion could appear just as a fortunate formal trick to obtain complex numbers behaviour for 1D mechanics and therefore not necessary so fundamental. Interest by the science community seems to have been mild, possibly also due to a later generalisation to 2D which, as should be expected, causes too many spurious processes to appear, and due to the apparent lack of insight brought to the problems of the probabilistic interpretation, of the quantum measuring process and quantum interference for multiple particles quantum systems. A "3+1" version of the model has also been more recently published, whereby a 1D wave function depends on 3D momentum, possibly providing some additional hindsight towards a full 3D extension - although this is currently unclear to me.

Whatever the others' judgement about Ord process, a number of considerations required at least some of my attention. (i) The first consideration was that the underlying dynamics used in Ord process, requires taking the continuum limit modulo 8 iterations of the dynamics (4 iterations, time two signs). This allows an elimination of the granularity of the underlying multi component process without loosing its continuum implications. This is interesting, among other reasons, because it suggests that the Maslov index is a bridging number required close to inversion points to recover the granularity of the underlying multi component dynamics without which the matching between the two sides of the inversion point can not be reconciled in the macroscopic limit; it was then the first time I could find a possible credible intuitive physical meaning of the Maslov index in the study of the semiclassical limit. (ii) The second consideration was that the Ord diffusion process implies an exchange between the diffused particles and the underlying rigid network. If one tries to model more precisely the dynamics, such an exchange can be more easily understood as a swap of diffused particles between the multicomponent gas an the network. Which in turn requires the rigid network to be built of similar entities, so to be able to undergo such exchange. Furthermore if one tries to model such exchange process, it is also necessary to admit that the infinite rigidity network used by Ord must be an idealisation. Therefore a more complete theory requires network structure to be discussed and any network variation or deformation to be modelled. This might provide for a simple conceptual reason showing why the localisation of energy (diffused light-like particles moving between nodes of a network) into quantum particles (quantum wave function) should cause persistent gravity like waves (waves of deformation of the underlying network). (iii) The third consideration was that the Ord process requires atypical assumptions about the dynamics, but it is a classical Newtonian process otherwise. Nevertheless it allows extracting the Dirac equation in one dimension, therefore suggesting that, at quantum level, special relativity can be reproduced from classical mechanics as an effective theory within the system where the underlying network is at rest. But then how is it possible that one can apply a Lorentz transformation, even in 1D and then find again essentially the same Newtonian process? There is either an inconsistency or there must be a way to extract also Lorentz invariance, and therefore special relativity in full, in some formal limit. But this in turn requires a model of the measuring processes underlying special relativity analysis and therefore a model of light absorption and emission by the multicomponent diffusion systems. Which in turn requires also a model of the photon within such a framework. This seemed way too ambitious, including because it does require a return to an underlying Newtonian dynamics, although with new matter structure considerations and, as discussed below, with electric and spin interactions only through direct contact.

With an out of the box view of the probabilistic interpretation and an intuition of the conservative imaginary viscosity features of quantum mechanics, the basic ideas behind Ord modelling seemed natural, although improvement and additional structure or ideas were required, particularly with respect to the justification of the dynamics, to the nature of the entities involved in the diffusion process and to the underlying network. The simplest picture to consider would be to use for the diffusion particles any image that contains handedness and charge of some kind. As these need to be exchanged with the underlying network, which in turn needs to be build of entities which connect points in some way, the simplest image to use is that of coiled strings (springs or left/right helices) with two handedness and two charges (2 binary properties, 4 basic strings). Charge would have to be understood as some form of plus and minus, either abstract as done in electromagnetism or modelled as two different linking structures along the string, as would be e.g. two different strands of DNA - each with their two handedness. But charge would be relevant here only for contact interaction and not for interaction at a distance. In the 1D case only one property is needed (parity in Ord papers), as the other property can be provided by the direction of movement. Nevertheless in higher dimensions, one would expect the 4 basic strings to be required to achieve the complex type rotation of Ord dynamics, which is required to get quantum wave function behaviour without Wick rotation, while the three directions of motion should be a cinematic input. Therefore it would be clear why Ord 2D generalisation causes spurious modes, using 3 binary properties rather than 2 (two directions of motion plus parity, rather than charge and handedness) while failing to separate intrinsic properties and kinematics.

*6) Splitting the problem: can it be done?*

Armed with these concepts different areas of modelling could be identified, each to be pursued independently at first, then each part to be used to try to put together a prototype model and if necessary each conclusion to be adjusted to allow a coherent synthesis.

*6.a) An electron instabilities inspired 3D multi-component diffusion model within a diamond network?*

By analogy with electron instabilities, a couple dynamics can be considered, whereby an underlying network would play an active role in establishing the coherence of a wave function that measures relevant parameters of couples' distribution. One can then consider quasi one-dimensional strings, coiled so to form springs, with two handedness and two charges. Charges can be intuitively thought in analogy of nucleic acid, i.e. reflecting some structure along the one-dimensional string, to be considered quasi one-dimensional, i.e. charge interaction is by contact and possibly mechanical in some way, rather that at a distance through some transmission mechanism. Although a charge internal structure cannot be modelled with the ideas hereby discussed, it is useful to keep in mind an image of quasi 1D objects, rather than idealised 1D objects with nil measure lateral width. Handedness can be thought as set by the shape, but also as spinning provided in principle the spinning is consistent to the shape. Allowing strings to break into strands (segments of strings) does further provide modelling flexibility and further reinforces the conceptual analogy with genomics (e.g. quarks as network knotted genes like structures). Strings can then be thought as not subject to network action to the point of obtaining some form of effective localisation, while strands (parts of strings) are obtained through some localisation mechanism which breaks the un-localised strings down. Un-localised strings are then best thought as existing mostly as pairs, possibly mainly of opposite charge and same handedness to provide a modelling basis for light quanta.

By common sense based on electron instabilities physics, one can then set a dynamics as follows. Strings (strands) can be coupled if handedness is opposite, charge is opposite, or both (a double helix if handedness is the same). The case of opposite charge but same handedness is a (virtual) photon like couple: for the sake of this high level description I will count one type, as the two handedness options are equivalent from a charge point of view. The case of same charge and opposite handedness is a Cooper-like couple: two types as two charges. The case of opposite charge and opposite handedness is a charge-density-wave-like couple: one type. As exchange has to take place with an underlying network, the most natural way to model the network would be with nodes set in space with a diamond structure, so to get isotropy of space in the macroscopic limit. Each node would then be connected to four other nodes with four strings, at steady state two positive and two negative, possibly all of the same chirality. Therefore building a quadrupolar network, possibly "spin 2" chiral in its steady state, where the nodes are the inter-knotting of four strings[6]. Spinning and wiggling (frequency) of the free strings or couples would be required to allow stability of the string configuration with itself and with the surrounding network. Upon collision with a node, a couple can then come out unchanged or modified through exchange with the quadrupolar node of one of the strings in the couple. This will temporarily modify the structure of the network, which will try to release back the spurious string to the next couple colliding. The temporary deformation and any angular momentum exchange will generate quadrupolar torsion waves and breathing modes across the diamond string network. As there are four strings building the couples and four types of stable couples, after any four collisions that modify the couple structure the couple will have gone again through the same starting configuration. If we assume a rigid diamond structure for nodes of the underlying network, this provides the 3D equivalent to Ord process in 1D. Where we to count separately the two handedness as well, an eight-periodicity process might be required.

The first formal problem is then to use such modelling principles to build a 3D version of Ord single particle process. Although the quantum particle in this case will not be just the result of the strings-couples-strands gas, but also of the composite contribution of the quadrupolar network. As the underlying network would then be common to multiple particles, it can be a mean for two-particle-interference similar to phonons creating correlation among the two parts of a Cooper pair. It should be possible for both particles clouds and network deformation to be superposed and act independently because of the hidden linearity of the multi diffusion process. Was such modelling possible, different spinor representations would then turn out to be different ways to count the components of the dynamics, i.e. single strings, couples, uncoupled strings two by two, mixed ways, other. Neutrino modelling would probably require only lateral diffusion and exact balance between plus and minus charges, therefore no or only effective longitudinal mass (and possibly small lateral mass). Amitaba Sen view in the nineteen sixties of the photon as two neutrinos might be incorrect, but possibly not so far off from just a counting point of view.

---

[6] Any two-dimensional version would then require a graphene like network structure, as undulating quasi 2D planes in a diamond network are projected on the plane as an hexagonal network: possibly a consideration if one wanted to use similar conceptual ideas to review the microscopic mechanisms underlying quantum hall effect in graphene. This embedded quasi-2D structure in the 3D network would also provide a microscopic basis for the quasi-2D decomposition of the gravitational force.

In principle, modelling in order to obtain known physics should provide elements to identify appropriate foundation assumptions and not vice versa. In addition there might be here a general vision that should allow bringing in a reasonable level of judgement at each step of such search. The vision is that there is here a kinematics and dynamics that, from a modelling point of view, is closer to soft matter and genomics. This is not just because of the photon simplified model provided part above and part below. But also e.g. because one can easily introduce configurations where the underlying network is saturated with charges of one type and the cloud of multicomponent multi string gas carries double the net charge and both travel coherently. This could give a meaning to the complex conjugate wave function. Furthermore one can easily introduce reversal of the underlying network charge polarisation at a boundary similar to cell or cell nucleus membrane boundaries. Knotting could help model hadrons and heavier mass excitations. Dynamics will be in principle not too dissimilar from modelling linked and unlinked polymers with solvent. If such a model can be build and turns out to have some relevance for the actual physics, then it would also be clearer why the spinor representation of the gravitational field is the most natural one: it provides a more descriptive formalisation of its microstructure (4 strings nodes connectivity, 4 nodes unit cell, although redundant because of counting constraints).

*6.b) Modelling light-matter interaction to get special relativity in the macroscopic limit?*

Michelson famous experiment showed that the speed of light in vacuum, as measured within the reference frame of the emitting source, is always the same, therefore showing that ether cannot be a medium for the propagation of light, a result later incorrectly extrapolated a bit freely as a proof that ether does not exists in whatever sense. From such a result to the Einstein hypothesis of a constant speed of light when measured in any inertial frame, weather the one where the light source is at rest or not, there is quite a leap. It is difficult to imagine that such an intellectual leap could be done without some essential additional idea. In fact the real motivation of Einstein hypothesis does seem to arise from the need to find a single new assumption allowing the derivation of the Lorentz transformations (Poincaré had clearly identified the need for a single hypothesis), rather than the pre-existing chaotic additional ad hoc assumption for each newly discovered phenomena in the study of the electrodynamics of moving bodies. Therefore Einstein checked the reverse logic, i.e. he assumed that it is not electromagnetism the effective theory (ether as a light propagation medium being one of the possible microscopic missing elements), but rather it is Newtonian mechanics to be an effective theory. Einstein approach, with its implications, was possibly easier at a time when time zones had not been defined and even close cities had official times often with half an hour difference, with the consequence of frequent train timetables mess. Einstein assumption is so simple and powerful and rightly celebrated. But it is also a barrier to fully transparent physical intuition, just as quantum mechanics, despite it works so very well. It is not so simple to see real grounds to revisit such approach today, although many have done so because of the abstraction and consequent lack of intuition caused by the accepted approach.

If we were to start just from Michelson result, without any additional element, we would have to consider emission theories. Such theories are based on the assumption that any light is emitted at a speed *c* in the reference system of the emitter, whatever the relative speed with respect to the reference system where the measure is being performed. But without any additional idea, such theories are fundamentally incompatible with Lorentz invariance, more or less at any scale, and therefore they are incompatible with the Maxwell equations. Therefore either the emission assumption is wrong (and therefore the special relativity speed of light postulate is the only possible viable option to derive Lorentz transformations) or there must be another conceptual ingredient at play that allows the derivation of Lorentz transformations, at least effectively within the appropriate scale domain. If Lorentz transformations were to be an effective formula, then the space transformation could possibly be looked at more naturally as a balance of viscosity i.e. action per unit mass (eliminate denominators), whereby the time transformation would be a balance of lengths (same: eliminate denominators).

It should be noted that there have been a number of speculative Plank scale discussions about breakdown of Lorentz invariance, which would seem to be triggered by the development of modern mathematical tools allowing building some extension of a theory, rather than for a openly stated specific reason driven by physical intuition. The most interesting one in my opinion is the use of deformation algebras (quantum groups) to obtain a deformed Lorentz invariance. Although this approach is once again triggered by the existence of the appropriate formal tools, it implies a physical intuition not far from the one discussed here. Lorentz invariance breaking to date tends to be tested through far source radiation measures. But these type of tests are unlikely to lead to a result, as absorption and reemission of radiation through not so empty inter-galaxy space makes the tests close to irrelevant: there is hardly a very long length through which light travels without ever been absorbed and reemitted, so the real distance over which the test is done is only the maximum length between an absorption and the next and not the radiation source distance to earth. But ultra fast lasers could allow testing a bit more at lab length scales.

My own reason to revisit special relativity is connected to the fact that based on the physical picture described in this paper, one could actually envisage possible mechanisms which could provide an additional ingredient to emission theories so to try to obtain Lorentz invariance as a effective by-product. Such mechanisms are ideally connected to the measures that must be operationally performed in order to derive special relativity. This would make sense if we remember that special relativity is a theory about empirical operational implications of measuring procedures, just like quantum mechanics. The relevant measuring processes that need to be analysed to formalise special relativity require light emission and absorption. Within the framework discussed under point (6.a) above, light quanta would have to be modelled as a double helix of plus and minus strings, with same handedness. Such double helix would need to wiggle its way through the underlying quadrupolar string network, with the wiggling giving the frequency and the step of the helix a measure of Plank constant. Absorption would then be a process through which the double helix is split in shorter strands, similar to how DNA or RNA would split e.g. within a cell core when injected, the equivalent of the cell being here the quantum particle, to be intended as the combination of the cloud of multicomponent strings/strands and the underlying quadrupolar network configuration, which could include e.g. polarisation or reversal of configurations. The broken strands of the photon double helix would then mix up to the strands contributing to the quantum particle wave function. Virtual photons would then be (partly) recomposed double helices that have not been emitted yet out of the particle cloud. And emission would be a recomposing process resulting in a full double helix strand to leave the quantum particle mixture. There would then be a time required for such process to take place. Such processes would take place at emission and, in reverse, at absorption. Therefore, provided the times required are the same, there would be hardly a way to see any influence in the observed time of flight of the photon. But as soon as the time required for the photon emission and the photon absorption is different, such difference would impact the time of flight. Such time would normally depend on the amount of string length and its density (packaging), with higher packaging at comparable speeds possibly requiring longer absorption time. The first parameter is independent from the relative state of motion of source and absorbing object, but the second is not and depends on the relative speed. A functional dependency that would reproduce the Lorentz transformations could then be searched for. But although one can in principle reproduce either the Lorentz contraction or the time boost, the mathematical requirements to obtain one or the other are somehow in conflict. This fact was implicitly well known and discussed pre special relativity, and possibly is one of the most important reasons for the success of special relativity, which through mathematical abstraction identify the Einstein postulate on light speed and bypasses the matter. But an approximate derivation of both contraction and time boost should in principle be possible highlighting the physical ground for a quantum group deformation of Lorentz invariance. Because of limited absorption free length of travel and because of a single starting point (big bang), any excess of speed over light speed in vacuum would be capped. Because of the acceleration of the expansion of the universe there should be none or very limited opportunity to generate extreme cases exceptions for such a cap. This type of modelling ideas might also allow revisiting the superluminal debate about phase vs. group velocities or the study of ultraslow light where macroscopic localisation of light would be the macroscopic quantum phenomena mirroring microscopic quantum light absorption. The recent announcement of up to 7 km/sec excess over light speed for the speed of transmission of neutrinos from Geneva to Gran Sasso might be related to the high speed of mesons generating neutrinos at Geneva neutrino factory and clean transmission of neutrino over one of the longest length ever tested. The fact that neutrinos arte hardly absorbed would be the reason for being able to observe super luminal speed in this case. Until recently it was impossible to have a good measure of the time of emission of neutrinos, therefore not being able to see such superluminal speed. But neutrino factories have changed this and emission time can be known more accurately. Therefore more superluminal speed announcements from neutrino factory experiments should be expected in the near future.

*6.c) Modelling of the underlying network deformations: merging gravity, Higgs, cosmological constant?*

The underlying quadrupolar network would be subject to several dynamical changes. Strings can be substituted through collisions generating distortion and torsion waves, collision without replacement would also generate distortion and torsion, compression or other forms of breathing modes can take place, as well as movement, network breaking an recomposing, knotting, defects, combinations of all. Spin waves like excitations could be gravitational waves, whereby due to the three fold interwoven planar structure of the diamond lattice they would be naturally decomposed in quasi-2D waves. Because coiled strings are the core building block here, a spin nematic type of order would be at play, suggesting that the metric field is a spin nematic order parameter, not far off some readings of the general relativity formalism. The condensation of such an order parameter being the result of free strings forming the underlying network. Possibly helped at staying together and under strain by the universe expansion acceleration. Because of the chirality of Ashtekar variables (original version) which can be set as opposite to the chirality of the electroweak theory, one would be tempted to model the steady state quadrupolar network with only one given handedness, whereby fields of strands couples have both handedness strands, but with the opposite chirality well in excess. Right left specialisation would be a driver of network versus string gas separation and condensation. Configurations whereby the underlying network has reversed chirality and so does have the strand gas would be the basis for antimatter modelling, which would clarify in what sense antimatter storage can be thought as a fuel production. Compression or polarisation modes of the network could correspond to Higgs degrees of freedom (holons as opposed to spin wave spinons). Average concentration levels of network components could allow to model a medium-scale local (or global) gravitational constant, which being also contributing to particles and selected other fields excitation would be a proxy of zero field energy in many cases.

Assuming by common sense a limit in the stretching of the network before tearing apart locally, possibly measured by Plank constant (coil size related), speed of light in vacuum (free spring emission speed) and spring tension constant (which should be related to the gravitational constant) one would then expect a gravitational force correction at very large distances, as the network would tear apart or not be stretchable beyond a given low density. If there is no tearing, the gravitational force would be stronger than Newton law at a given central mass and beyond a given distance, similar to the semi-empirical formulas identified by M.Milgrom in the study of galaxy rotation curves. In case of tearing, it would be lower. Similarly at very short distances, one would expect a gravitational force that is lower as there should be a maximum compression, also measured by the same constants. Else breakdown of the network, which implies even lower force. The very long and very short distance corrections would mirror the correction versus classical mechanics of the multi-diffusion process, which prevents collapse or total dispersion of the quantum particle. It might also contribute in thinking about so called dark matter and dark energy gravitational force discrepancies and reduce the role or the options about the nature of WIMPS required to explain such discrepancies. 2D of solar system and alignment of most planetary spins would be possibly a consequence of the quasi-2D decomposition of network dynamics at planetary system formation.

A useful almost poetic image, with its usual limitations, to think about the underlying string network is to think at a 3D spider web, with water on the web as a proxy of the condensation of free strings gas drops. As a spider stands on the four back legs and knits with the four front legs to do its web – and we also knit with four double pointed needles to do socks, whereby two needles are needed to hold and two to pull, so to be able to hold and model in 3D – so nature uses the four link of the underlying network (possibly right handed in its free version) to hold up the four types of free strings (possibly left handed in their free version)[7]. Energy condensation (matter) then holds the web under tension. The global web is held under tension by the big bang drift and the accelerating expansion of the universe, which also prevents extreme conditions and enables the formation of the extremely fragile environments, which in turn allow life to form and survive.

This conceptual framework could be used to revisit some of the most popular unification theories: each of them could just be simplifying the microscopic kinematics and dynamics by making effective assumptions. String theory uses the right quasi dimensionality of basic entities and takes in the general framework of quantum field theory, with supersymmetry fields taking into account collective modes as if they were separate disconnected degrees of freedom, including possibly of the underlying network. But no explicit signature of the underlying network or of its links to particles fields is visible and extra dimensionality covers for that but as independent degrees of freedom (similar to the intuition given to the Kaluza-Klein extra dimension by Einstein, according to Pais). Lack of details of a springs-strands-couples dynamics is possibly balanced by quantum field theory and regularisation technical details. In loop quantum gravity one starts from the Ashtekar variables where the complexification takes into account in the continuum limit the condensation of the underlying network and of the concurrent condensation of the massive particles combined cloud of string gas and network variation; loop quantisation on the top then provides the equivalent quantisation procedure partially bypassing the details of the microstructure, which appear in a less explicit way in the structure of the quantized theory. Therefore spin networks mirror the underlying string network and quantum groups are required for the regularisation of measurable quantities reflecting the non-zero size of the string coil width[8]. Non-commutative geometry frameworks are about trying to represent with separate, dimensionally disconnected degrees of freedom the macroscopic limit and some of the local details of microscopic discrete dynamics. Scale relativity is about seeking a representation whereby there is separation between the steady state microstructure of the underlying network (scale invariant above an appropriate renormalisation scale) and its breathing mode. Regge calculus models reflect the underlying network geometry. Each of these approaches then would be acceptable in some regime, whereby the fact that an approximation is taken as an assumption does result in asymptotic series for the relevant outputs[9]. Lisi theory of everything is maybe about the full symmetries of the underlying dynamics of strings and network, including collective modes. And graviweak unification frameworks seem to be the correct base on which hadron physics could be further clarified, e.g. colour as knotting and flavour as quasi-low dimensionality.

---

[7] The right and left specialisation is both inspired by the Ashtekar variables discovery, by Penrose comments in his book *The Road to Reality* (2004) and by the traditional Jewish view of the symbolism of right and left specialisation in the human body (i.e. in biological systems) and its relation to the way God acts in the world, allegedly more explicitly so at the time of creation.

[8] Quantum groups are clearly neither "quantum" nor "groups". They are algebras resulting from a deformation of a starting symmetry group, such deformation taking into account the non-zero size (core rigidity) of the entity providing the center of the symmetry. As in most relevant cases such core rigidity is by a very small size in relative terms, one can think at quasi-groups. As the first cases of important practical interest had the Plank constant as a measure of such core rigidity, many do call them quantum groups. So the simplified intuition in the case here discussed is that the connecting lines on the spin network are to be considered as coiled strings (springs). Therefore the finite transversal size of the coil would require the relevant symmetry group around such direction to be deformed into a relevant quantum group. If such deformation is overlooked, infinities do arise in spin foam graphs calculations.

*7) Digging among original ideas to test the recipe: possible bridges among generations?*

During the past four years I have received as unexpected presents many books among the classics in physics. I could then read in English, mostly while standing in the London underground, some of the original writings of Kepler, Galilee, Newton, Carnot, Drude, Faraday, Maxwell, Hertz, Lorentz, Poincaré, Plank, Einstein, De Broglie, Born, Heisenberg. Some reading required looking further back for other and sometimes older classics, some of which are now available online. Some topics have provided further guidance and intuition for my search: (i) covering the two revolutions of thoughts about inertia, the first led by Galilee and Newton, the second led by Einstein; (ii) covering the background of the initial ideas at the roots of quantum mechanics. To look forward as far as possible, it eventually helps to look back.

*7.a) Source of inertia: divine, then an intrinsic force, then a dynamical property of curved space. Why not a dynamical property of the underlying microstructure of quantum fields?*

Avicenna, Maimonides (12$^{th}$ century) and Aquinas (13$^{th}$ century) called on Aristotelian physics to build a cosmological argument of the existence of God, possibly making a good faith conceptual error. In part it might have been the result of an idealistic and/or manipulative way to reconcile science and faith of their time, both having very heavy imperfections, just as today. A simple summary of the matter at hand: it is an implied assumption that empty space is an abstraction and however empty, space always is filled with something. Therefore however small, this something would provide some resistance, just like air for a bullet, which would eventually stop an object travelling at a finite speed if one waits for a long enough time. Therefore, if something continues to move in a substantially unchanged way for a very long time, without an apparent force maintaining its forward momentum, it must then be pushed by God or heavenly angels. Therefore the movement of heavenly bodies was called upon as a proof of God existence, including for the fact that an incorrect assumption was made about the persistence and unchanged nature of their movement.

It is interesting to remark that the starting assumption about empty space is correct, at least in some sense, according to modern physics. Regarding the deduction, for the faithful, God is ultimately the source of all things and therefore the idea of God as a source of inertia does not require much analysis and is rather a high level explanation based on faith. On the opposite, requiring that God is directly pushing each object so to maintain its state of motion is a bit arrogant in a way, even more for the faithful. Furthermore, the acceptance of Aristotelian physics at that time led to the additional abstraction that quasi-empty space is actually empty and that even in that case a force was to be required to maintain the same status of motion, therefore making the *Prima Causa* proof of existence apparently stronger. Such *Prima Causa* theory is just like saying that the source of inertia is direct divine intervention. From the later Galilee experimental work, the Aristotelian view is known to be empirically incorrect, therefore making the logic of the theologians a bit silly and the rationalistic conclusions about the existence of God a bit awkward, being based on incorrect physical principles.

When then Newton formalised and sealed the idea of inertia as an intrinsic force or resistance of all what has mass, a *vis inertiae*, he then used the concept as a key ingredient to build a derivation of Kepler laws from Hooke intuitive law of gravitation, allowing to show that what is below (weight) is the same of what is above (celestial motion), in line with the Hermetic tradition among alchemist, so dear to Newton.[10] The alchemist background of Newton is quite relevant as it most probably influenced him at

---

[9] Asymptotic series for a given explicitly known or unknown function are generated by extending to the full domain being considered a simplifying assumption valid at least with good approximation only within a limited but crucial domain – crucial in terms of numerical contribution. This makes it possible for approximations to be more precise when using the first few terms of the series, but diverging on from a given term on. Resummation techniques can at times be used to try to find the original function. The best guidance to look for some form of renormalization at each order is generally some knowledge of where the starting assumption is not fully correct. For the case here discussed, a multicomponent dynamics within a dynamical quadrupolar network is a very tough model to formalize or calculate, but it seems to be conceptually clarifying where the other theories might be extending assumptions not globally valid.

[10] Seen form a scientist point of view, the Aristotelian physics based deduction so dear to the middle age leading analytical theologians of all three monotheistic religions was driven by the search for some "microscopic" description of inertia: lacking more advanced theoretical ideas God was directly called upon. This was then institutionalised by the rejection of the experimental method supremacy on pure logic, which being more easily prone to manipulation is historically preferred by theocratic unchecked and unbalanced authorities. The conclusion about the existence of God, as the less analytical theologian knew at all times, is quite likely a matter of faith. In my very personal opinion faith can be found in three key forms: (i) in its stronger form is driven by deep intimate personal experience, therefore not reproducible in a controlled way, which makes the analytical theologian search of an analytical proof through cosmology incorrect and the empirical definition of "believing in God" subject to uncertainty, as each one of us is our own and only sample, we are all different and have different experiences – the confirmation of existence is therefore intuitively inferred by comparing notes and searching for common grounds, very difficult and possibly impossible to conceptualise, although today's helped by new tools like social networks; (ii) in its weaker form is driven by

considering ether ideas. In fact, the allegedly Egyptian-Greek ancient Hermetic[11] based tradition of alchemy included the belief in the existence of an ever all pervading substance which allowed effectively anything to be transformed in anything else. This was really an evolution of the concepts related to the Egyptian divinity Shu, or later Greek divinity Aether, which was seen as a connection mean between the earth and the sky, the sky being a container for all what can be seen from earth by looking up. Therefore Aether, and later the ether, was the key link between what is below and what is above.[12] Based on such concepts many alchemists believed in the existence of such all-pervading ether, which implied in principle that also changing vulgar metal into gold would have been possible. This change is now known to be conceptually possible, although not through chemical changes as the alchemists wrongly believed, but rather through very extreme nuclear reactions quite difficult to produce, certainly on planet earth, but also around the more nuclear active universe.

Whatever we might think about the source and evolution of this type of ideas, it seems that Newton believed in the existence of such omnipresent medium and this might have encouraged his finding of a link between what is above (celestial mechanics) and what is below (weight). But to reconcile the idea with his law of inertia and not to fall back into a *Prima Causa* explanation, he had to assume that such medium, if it existed, had to be resistance free. Because of its properties, it would have been then a medium that could mediate gravitational force, therefore resolving the logical inconsistency of his gravitation theory that at face value, as an effective theory, would require action at a distance.

It is then less surprising that related ideas do appear in Einstein essays, at a time when he was trying to develop further the physical intuition and meaning around his theory of the gravitational field. In fact, quoting Albert Einstein concluding remarks from a University of Leyden talk delivered on 5 May 1920 (from A.Einstein, Sidelights on Relativity, translated by J.B.Jessery & W.Perrett, 1922): *"we may say that according to the general theory of relativity space is endowed with physical qualities; in this sense, therefore, there exists an ether. According to the general theory of relativity space without ether is unthinkable; for in such space there not only would be no propagation of light, but also no possibility of existence for standards of space and time (measuring-rods and clocks), nor therefore any space-time intervals in the physical sense. But this ether may not be thought of as endowed with the quality characteristic of ponderable media, as consisting of parts which may be tracked through time. The idea of motion may not be applied to it."* This concept of "ether" by Einstein is in a way a sophistication of the one by Newton and there are differences, including e.g. that for Einstein light can not exists without a gravitational field because an electromagnetic field, even very small, would always generate a gravitational field according to general relativity - while for Newton it could be conceived that light can exist without ether, which in turn does not necessarily interact with light all the times. Einstein concept of the ether was possibly closer to the empty space of Aristotle, although with opposite inertial properties.

Having said that, it is quite intriguing that the framework discussed in this paper requires an underlying network to mediate quantum diffusion and interference, enable energy localisation, mediate gravity as a spin wave, provide local empty space holon like energy concentration (vacuum energy / Higgs), etc. This is somehow close to Newton ideas, and possibly only partially different to Einstein tentative concepts. It might include both of them in a way, but with some different conclusions, the fist of which is that such "ether" is not in general a medium, except possibly for gravity waves, but contributes in a different way to most fields and influences all fields' evolution, at times like the Higgs field concept, at

---

circumstantial considerations or intuitive fear, which is possibly what brought to the subject the analytical theologians; and (iii) in its pathologic form is driven only by compliance requirements set by theocratic-like power, therefore about the relationship between religion and political power rather than about spirituality and/or belief – it is this last kind of faith which has been used to enforce an incorrect, non-experimental based knowledge of the law of inertia for many centuries.

[11] Hermes being the Greek name for the mythical Egyptian king, priest and philosopher (three times great, or Trismegistus), then raised to divinity by ancient Egyptians, and later by ancient Greeks.

[12] This concept of connection between earth and heaven, material and spiritual, was a big deal in a quasi-theocratic Christian society, as religious institutions probably saw it as a great danger likely to lead less educated Christians back to paganism and therefore related ideas were at times branded with a status of heresy. Although much has changed for each of Christianity different forms today, Christian institutions were initially structured with the belief that, at least in certain areas, material and spiritual interaction is dangerous and a full conceptual Greek philosophy inspired duality was required for safety reasons, including enforcing a stringent degree of separation between the material and the spiritual in certain cases. Trying to put them back together was then undermining some of the foundations of the whole system and therefore forbidden. The concept had then to be replaced with a spiritual only concept - the *logos* of Heraclitus, and later of John the Evangelist, providing a perfect option - therefore avoiding the perceived dangerous coexistence of material and spiritual (the material side was addressed separately with the physicality of Jesus), but influencing in a biased way and limiting scientific thought for so long. It is ironic that modern information theory and linguistics certainly would be able to represent the string network, here suggested as a key ingredient of quantum reengineering, in a way that could allow us to abstract its dynamics as a speech, i.e. a *logos* in the literal meaning of the term. Cabalistic influence on Jewish daily prayer also refers to God continuous involvement with creation as a form of talk or word, as it is an active action, but mechanically almost effortless. The interplay between strings network on one side and diffusive strings on the other side discussed in this paper does provide an *a posteriori* interpretation of the Tohu and Vohu of the Genesis, which in turn shows how the knotting link between physical and "spiritual" is again a driving concept behind the theoretical framework here discussed. It should become also clearer what is the connection between the Tohu and Vohu of the Genesis and similar philosophically more developed concepts in other religions, such as e.g. Ying vs. Yang or Tao symbolism.

times like in the composite fermions contribution by the field contributing the flux quantum. This tentative conceptual bridge between old ideas and ideas here sketched does single out a fundamental property which any prototype model of a string network for quantum reengineering would need to have: it has to be "resistance free" in the sense that its own adaptation and variations must allow particle configurations to travel through it free of change of their status of motion, in a soliton like fashion, so to allow the law of inertia to be obtained in the macroscopic limit. Therefore components of the network must have ultra fast adaptation dynamics. This probably requires modeling ideas more ordinary in soft matter physics and biophysics. As a by product, the principle of equivalence would then have an intuitive geometrical base: it does not matter if a particle configuration is moving through the network under the effect of a nematic-like spin wave (gravity) or out of action of another force; all what matters is the relative dynamics between the (composite) particle configuration and the network itself. If a difference was to be found (i.e. some form of weakened equivalence) it must be searched in the mechanism of transmission of the spin distortions as it might result in different smaller relative configurations in different cases. It is also interesting to remark that such strong network would represent the rope in Newton *Principia* analogy between planet gravitation and hand made rotation of a stone attached to a rope: the rope resistance provides the centripetal force which balances the centrifugal component of the stone status of motion and allows orbital motion; similarly the string network does provide a physical connection, although it is not just by simple resistance that the centripetal force is generated, but by the combination of knotting (keeping the strings of the network together, i.e. resistance or tension) and spin waves on the string network (dynamical torsion).

*7.b) Shaking versus mixing: could macroscopic irreversibility be affected by quantum reduction?*

Planck was intrigued by two features of the empirical laws of black body radiation, which became clear from the related XIX century experimental discoveries: (i) it does not depend upon the black body being considered; and (ii) it has a quite simple functional form. The first feature implies the universality of the key macroscopic constant used to describe the distribution of radiation, a constant that would have normally been expected to be a function of the substance, as in other statistical mechanics problems. The second feature implies simplicity of the underlying phenomena. He then searched for a way to extend the area of application of statistical mechanics to electromagnetism, effectively searching for a unified treatment of electromagnetism and thermodynamics. Therefore he was looking for one of the two missing links to conceptually unify the three broad areas of physics theory of his time, the other one being the link between mechanics and electromagnetism, which was shortly after provided by special relativity. At the time Boltzmann had already found the first of the links, the one between thermodynamics and mechanics. This three-way links program was at the time a framework similar to today's unification program[13]. Using the simplest microscopic model (harmonic oscillators) for the vibrations modes of the molecules building the black body internal surface, Planck then looked for the statistical treatment of infrared absorption and emission that would result in the observed functional form of distribution of heat radiation. The universality of the constant in Wien displacement law implied then the universality of the constant involved in describing the microscopic process, which is today's known with Planck name. To bridge thermodynamics and electromagnetism one had then to find a Boltzmann constant for electrodynamics, whereby temperature is replaced by frequency. Therefore such a constant had to be an action constant. In Planck description, the black body molecules vibrational Raman modes allow the exchange of energy with the electromagnetic field and back, going through equilibrium configurations. It is then as if the black body is acting as a measuring instrument on itself through the electromagnetic field, enforcing an adiabatic process thanks to the macroscopic size of both the black body and its heat radiation. In the simple case of quantum measuring process on the opposite, the system being measured goes through a non-equilibrium transition and because of the rigidity (calibration) of the measuring system, it could be forced at time towards the less obvious configuration, provided other similar systems compensate for that. From the point of view of ergodic theory, it is as if the set of equal systems is shaken so that some of the samples do occupy configurations that would otherwise not be obtained within a given time. The question then arises if in traditional ergodic theory one should not explore the effect of a shaking effect of this kind, on the top of the usual mixing. And how this could allow temporal means to get closer to space means in a shorter time, in order to compare to measured data: the shaking by events similar to quantum measurement might speed up the mixing. Which might be an indirect signature of what hereby discussed.

---

[13] The cross feedback among the foundation ideas of each of these three areas of nineteenth century physics theory, generated by the search of unification, contains most of the seeds of the so called twentieth century physics revolutions. Mechanics required an atomistic statistical underlying description of thermodynamics. Thermodynamics then required the quantization of the electromagnetic energy exchanged with matter. Electrodynamics required a constant measured speed of light and therefore a modification of mechanics. But the cross influence is not really coherent and to date any gap has been bridged by quantum theory. Nevertheless, a tentative higher level of coherence could possibly not be achieved until one also (i) recognizes that even classical mechanics is a form of thermodynamics, (ii) accepts that phenomenological quantization "from above" does not close some of the inconsistencies and (iii) dares considering that a more advanced description of measurement processes does probably impact the understanding of special relativity as well.

*8) Checking other views of quantum mechanics: an echo from the more recent past?*

At least two checks were then required before validating any plan of work: a review of Bohm and Nelson ideas about quantum mechanics. Regarding Nelson quantum mechanics, it seems a posteriori to be effectively a statement about the net stochastic processes that are here proposed to underpin quantum mechanics equations, therefore allowing the separation of the classical flow from the "quantum potential" to be modeled with a single component stochastic process. It is missing the multicomponent underlying dynamics and the idea of a structured contributing underlying network, but it provides an extremely powerful technical tool for selected semiclassical analysis work, which is possibly the key reason why it has been so much celebrated. It also identifies the imaginary part of the complex semiclassical velocity as an osmotic velocity, therefore enabling some of the same intuition here discussed. The same type of intuition at macroscopic level has also been used e.g. by Feynman in his college course discussion of Ginzburg-Landau theory superconductivity, where the same mathematical term appears as a function of the superconducting complex order parameter.

Regarding Bohm causal interpretation I did read much about during my first couple of years of college, and I was under the impression that it was a formalization and evolution of De Broglie ideas which first encouraged and was later encouraged by the discovery of the Aharonov-Bohm effect. In fact Bohm framework does not lead in any way to multicomponent dynamics, nor it leads to underlying network interaction. Nevertheless, Bohm later in his life got interested in other fields as (i) physiology, with his holonomic brain model, and (ii) sociology, with his dialogue concept, which is really a precursor of social networks subgroup dynamics, where group cooperative creativity share and enhancement might be the social equivalent of a small endothermic shock. Because of the ideas he experimented in these other fields and how he drew parallels to physics, I felt that a further check might have been worthwhile, in order to see if any broader concepts had been raised at any stage. In June 2011 I could read for a couple of hours a few letters written between Einstein and Bhom, copies of which are kept at Birkbeck College Library, as part of The David Bohm Papers. From my reading of such correspondence it would appear that the causal interpretation by Bohm might be the result of a thought process aiming at a formalism that can be mathematically manipulated. Such thought process must have required a number of compromises, but it did generate initially from conceptual ideas possibly partly in line with this paper. It is further not surprising that Bohm did focus so much on the matter during a time of quasi isolation from the scientific community and limited knowledge of the local language, a time of excessively reduced means of expression: if you have no means to do focused research, which tends to be more rewarding, you have to choose a subject which can be studied with no means at all! I certainly felt sympathetic.

*A letter from David Bohm (in Sao Paulo) to Albert Einstein (in Princeton) on 3 Feb 1954, included the following (from The David Bohm Papers - reproduction authorized by Birkbeck College Library, as agent of the copyright holder):*

" As for my work I am beginning to think in a new direction. I am inclined now to think that the $\psi$ function represents a statistical property of matter. The general idea is that at a level more fundamental than that of quantum mechanics there is a field which satisfies causal laws. This field is however in a state of statistical fluctuation. These fluctuations are somehow described by the $\psi$ field. In this way, we could understand why the $\psi$ field must be expressed in a multi-dimensional configuration space (since statistical correlations require just such a space for their description). The relation of the underlying filed to the $\psi$ field should be analogous to the relation between Brownian motion and the underlying molecular motions that give rise to Brownian motion. This point of view tends to approach your idea that quantum theory is now "incomplete" (since it leaves out the deeper level causal law and treats only of the statistical laws appropriate at the present atomic level). However, as always, [I feel that a better clue to the deeper causal laws lies in a more detailed study of the atomic or sub-atomic levels. I feel doubtful that the deeper causal laws are clearly and unambiguously reflected at the macroscopic level] – although it is, of course, not entirely out of the question that a careful study of electricity and gravitation at the large scale will provide valuable clues as to the forms of causal laws operating at the subatomic level."

*On February 10, 1954 Albert Einstein reply to the above letter included the following (Albert Einstein Archives 8-041, reproduction authorized by the Albert Einstein Archives, Hebrew University of Jerusalem, as copyright holder):*

" I am glad that you are deeply immersed seeking an objective description of the phenomena and that you feel that the task is much more difficult as you felt hitherto. You should not be depressed by the enormity of the problem. If God has created the world his primary worry was certainly not to make its understanding easy for us. I feel it strongly since fifty years."[14]

---

[14] Einstein view is, not so surprisingly, possibly in line with the Jewish traditional reading of Genesis 3:24 - *And having driven out the man, He stationed at the east of the Garden of Eden the Cherubim and the flame of the ever-turning sword, to guard the way to the Tree of Life*. Which is understood, possibly metaphorically, as indicating the presence of agents of God at the entrance of the Garden of Eden who are guarding the access to the way to the tree of life (metaphorically understood as the full implied content of the five books of Moses or, more broadly, knowledge of the world - the second of the trees from which feeding requires first the right preparation or else tends to cause suffering). At the same time the same agents are making light to show the way to those who can come with the

The reading of this reply by Einstein did generate a degree of commotion to me, as in 1992, coming from a completely different angle, when I started to unfold the almost uncountable common sense questions generated by the problem at hand, I did to a certain degree get depressed by the enormity of the problem. But I do now feel that the path towards a prototype answer might be closer at hand and can be tested. The fact that to fully identify the path one has to give up on special relativity as an absolute theory, might be the reason why to date the De Broglie-Einstein-Schrödinger-Bohm approach has not led to a full set of key ingredients. In fact, how could Einstein revisit special relativity when he was spending so much time defending it, often from ill motivated attacks? And how could the others dare exploring a route potentially in conflict to the dearest product of Einstein intellectual creativity? But purely mathematical considerations – such as (i) dimensional reduction of variational methods which identify countable local variables together with comparison with viscosity through semiclassical formalism or (ii) that to be able to obtain a local quasi-complex structure for the quantum phase one needs a circular process and therefore the appropriate multicomponent diffusion – and other considerations, made with sound physics judgment, might now be showing a different possible way.

**Brainstorming about selected possible implications.**

*Brainstorming on neutrino factories potential and gravitational waves detection.*

If hadrons and heavy particles strong energy concentration can be modelled by knotting and underlying network polarisation, then radioactivity is the unfolding of a fragile or instable configuration. Neutrino matter interaction could then allow other possibly more structured forms of induced instabilities. Therefore neutrino factories could be able to discover new forms of induced and micro controlled nuclear decays. The principle is that, if neutrino beams have high intensity, unfolding of less unstable configurations might be possible. Hopefully not to include induced proton decay!

From the point of view of this paper, gravitational wave detection would best be obtained with the use of 2D antennas in orbit. For incoming waves, it would be preferable a fully planar detector on an orbit as stationary as possible against the fixed stars. For outgoing waves detection, a spherical detector on a geostationary orbit would probably be the preferred option. Outgoing waves would be earth generated, with low intensity and low noise if selected on the basis of planar correlation. A large scale review of earth crust and internal movements would allow taking a view on best option of detector geometry and location, just by using proxies of the Einstein bell example for gravitational wave generation. Incoming waves would be at times stronger (extraordinary astrophysics events) but only rarely and with higher noise. A simple analysis of realisable orbits versus directions having higher probability of receiving signals from a stronger gravitational wave generation event could allow taking a view on this type of speculative approach. In principle detection would be best based on quadrupolar spin 2 excitations as e.g. with some of the suggested d-wave superconductor detectors.

*Brainstorming about areas where similar modelling principles could be applied.*

(1) Neuron cells form a 3D network of quasi 1D building blocks whereby complex and at times 3D dimensional cross excitations are transferred along concurrent multiple paths. In addition glial cells interact with the neuron network in different ways, including affecting the neurotransmission, removing neuron cells and modifying the network during its early formation process. There could be then a different way to model neural activity, at least for larger areas as the brain, whereby exchanges within and between glial cells and neuron cells are seen as a multi component diffusion process in a network. This might allow a bridging between the Pribram-Bohm ideas about a hologram like mechanism for thought formation within the brain and the more traditional sequential views, by bringing in anomalous dynamics factors. This type of approach to brain physiology and thought formation modelling might further clarify the strength and weaknesses of Penrose like ideas about quantum features in the way the brain and mind interact: there might be some correct intuition, but the conceptual idea of where quantum like features come into play might not be correct. Rather than full quantum effects, just partial quantum features of an underlying multicomponent dynamics on a network might be at play. A strict quantum analogy might not apply and criticism about Penrose conceptual ideas might not be fully correct.

(2) A market transaction can be looked as the result of the matching of a buyer and a seller. Each matching can be initiated by a buyer or by a seller. Therefore there can be four types of prices: making price versus accepting price and buyers versus sellers. It is a couple matching dynamics among four types of prices, mediated by a market infrastructure (exchanges, trading lines connectivity either phone or electronic) connecting trading sides. When the resolution in the evolution of the price is comparable to the minimal size of price variation, typically in the study of the microstructure of financial markets so important for high frequency trading, some

---

appropriate intention. From this point of view the meaning of the etymology of the Hebrew word Cherubim is connected to the, sweet, intimate search for fundamental knowledge. Later in history, such conceptual themes have contributed e.g. to the symbolism attached to the search of the Graal.

(3) The development of the interactive Web between 1999 and 2004 has generated a media creative wave that is becoming a full 21$^{st}$ Century Renaissance. Its most visible expression, social networks, allow the study of social dynamics among a very large number of people, including clustering by interest, subjects of interest across subgroups and so on. They are the natural platforms for the new generation of communication and information sourcing and analysis. They have the full potential to become a key tool for opinion polls, premarketing, product engineering, political strategy setting, financial markets pre-shock monitoring, daily news broadcasting, daily ordinary communication, provided they can continue to give free new services to those who allow the network to own their opinions and information. They can be an instrument of dictatorial control or of democratic breaking of the domination of a few; of more mature consumerism or improper commercial behaviour; of more mature media or more manipulative propaganda; of improved sentimental education or extreme unchecked sexual experimentation - depending on how they will be used. They will probably require a regulator, at least for part of their activity, as possibly many other software solutions industries. It should be possible to model important aspects of dynamics on social networks with multi-diffusion processes, whereby diffusion components will be many and network topology highly variable. Full meaningful modelling would be impossible and it will be important to identify key network features and key information exchange to model a specific feature correctly. Phenomena similar to quantum coherence and de-coherence or interference might be possible e.g. among different subgroups, whereby information is not shared outside the group until some form of coherence breaks down. Although the booming of social networking culture is currently trending towards a new form of Gnosticism, it might actually be the result of an unconscious understanding (i) of the existence and quality of interconnections, both physical at the fundamental level and in our thought process at a physiological level, and (ii) of how at a social hierarchical level the same class of mechanism does allow to stabilise our so called consciousness (some net statistical thought process) in a steady state, therefore avoiding preliminary or degenerate forms of folly (full dispersion or de-coherence of a wave function like process). Social networks should then be considered at other levels as e.g. therapeutic (uncontrolled virtual group therapy – a lot of people do use them this way already) or for empirical research on psychology stability and transformation, including to and from pathological forms – i.e. for a broader out of the box study and review of the foundations of behavioural science. Faith experiences could actually be tentatively modelled one day through this type of analytical work: see also note 2 point (i). This might be possible because most strong faith experiences are related to (a) the fundamental breakdown of communication connectivity, as caused by too fast and too incompatible cross requirements on the individual, and (b) to the extreme condition of rebuilding such connections from scratch, which in turn provides a unique know how about the experience of ourselves - as unique samples, in principle not even closely reproducible in a lab - and of how such experience is interconnected to the interaction with others, which is at the foundation of human spirituality, personal identity and self esteem[15]. Conversely, excessive psychological conflicts or loss of faith tend to be caused by excessive constriction and abuse. Isolation from the abuser, higher level of isolation and then time might allow to reconcile conceptual conflicts through a renewed social and emotional connectivity, which can rebuild self esteem. The time for such processes to be more explicitly understood, hopefully also by those less prone to abuse and manipulation, might be closer.

**Closing remarks.**

Although many of the ideas here presented might turn out to be incorrect or incomplete, if anybody could dedicate enough time to search for a full prototype model along these lines, such search could provide a natural way to look at the relevant physics so to identify useful conceptual simplifications. Furthermore, a proper research project along the lines discussed in this paper would normally need to take place very close to experiment and applied physics, with the goal of identifying frontier measures to test or try to invalidate the conceptual framework hereby discussed. The possibly key driving element remains that both quantum mechanics AND special relativity are theories about operational implications of measuring procedures and there can be no full microscopic theory for one without achieving the same type of result for the other. There was possibly an element of folly or unchecked creativity in the search I here described and I could let curiosity go where it was natural to go and had fun. Hopefully it can amuse also those readers who don't get displaced by the jazzy style. It is an interesting intellectual exercise much for the chain of questions it raises, maybe even more than for the answers hereby sketched or presented as heuristics. It certainly helped keeping my intellect well alive during some difficult times. Ultimately I am just saying that there is maybe a way of counting and of choosing what to count that has only been partially identified to date and which in turn, if correct, would contribute to multiple areas of knowledge, just as any way and choice of counting does. At the broader level I am just saying that similar concepts do appear to connect a bit more clearly quantum physics to some fundamental areas of life and behavioural science.

---

[15] A review of the concept of spirituality, seen as a coherent state of body and mind and based on analogies to the physical ideas discussed here, could be discussed together with implications about tools to manage its risk and rewards. Including in order to compare, at least at a theoretical level, religious approaches as different risk-reward choices. And to compare different religious practices as development of different management tools. But this is quite a broad topic and possibly a complete separate conceptual exploration.

**Acknowledgements.**

Complex frequency work was done between 1990 and 1991 under the supervision of S.Graffi; small viscosity limit work between 1991 and 1992 under the supervision of C.Viterbo; photoemission work in 1993 under the supervision of G.Margaritondo and F.Gozzo; MPTBp X-Ray work between 1994 and 1996 under the supervision of J.P.Pouget.

In 1992 S.Fubini clarified to me the importance of macroscopic quantum phenomena and the background of his interest in anyon superconductivity and composite fermions. In 1993 R.Krikorian offered me a desk in his office for six months, which provided me with a lifeline and a base, allowing me to reassess my research interests. In 1993 M.Heritier gave me the opportunity to choose a research subject in condensed matter physics among a wide list, including allowing my full attendance to the classes for a post-graduate diploma in solid state physics. In 1996 M.Venturi gave me as a present his university books on biology and biochemistry. Kings College Maughan Library provided me free access on Sundays for personal research between 2003 and 2005. In 2004 T.Kibble provided astronomy and astrophysics references on arXiv. In 2005 P.Littlewood refreshed my memory about some aspects of BCS theory and Cooper pair formation, shared his views on Zhang SO(5) theory of superconductivity and antiferromagnetism in cuprates and usefully commented on requirements for any mathematical framework aiming at the construction of a quantum wave function. Interest in history of physics was stimulated through reading of articles provided by S.Bergia in 2005. Translations of selected milestones books in the history of physics were presents made separately to me by my now separated wife Isabelle and by J.Van Messel between 2006 and 2010. In 2011 Birkbeck College Library provided access for a couple of hours to letters between Einstein and Bohm, which are part of The David Bohm Papers. In September 2011, the Einstein Archives at Hebrew University checked the quotes and sources from A.Einstein, including if authorisation was required.

**Post scriptum: are special relativity and quantum mechanics a modern form of mysticism?**

In my opinion yes, they are a sophisticated quant modern form of mysticism, in the sense of deep correct-in-some-way intuition bypassing full logic understanding, which in turn, because of the synthesis achieved by the choice of postulates, result in ultra high theoretical precision. And this partially clarifies the initial unbalanced, abusive and unmeasured reactions to the theories, the best example being the German nazis ("Jewish science") and Soviet communist ("bourgeois science") attack and prohibition of the study of special relativity. It possibly also clarifies De Broglie, Schrödinger and (ironically) Einstein reaction to the Bohr and Born schools' approach to quantum mechanics. But in my opinion such modern mysticism can be transcended today, in the sense of filling in part of the bypassed logic understanding, thanks to an amazing advance in scientific knowledge when compared to when the theories have been first formulated. And so many have initiated such transition and explored routes. This paper intends to try a broad horizontal approach, although only conceptual, so to set a research plan for those who want and can afford to pursue such a speculative path.

**Selected bibliography.**